\documentclass[
    ,final            
  ]
  {aipproc}
\usepackage{amsmath,amssymb}
\layoutstyle{8x11double}

\begin{document}

\title{$J/\psi$ production with NRQCD: \\ HERA, Tevatron, RHIC and LHC}


\author{Mathias Butenschoen\footnote{Speaker} }{
  address={{II.} Institut f\"ur Theoretische Physik, Universit\"at Hamburg,
Luruper Chaussee 149, 22761 Hamburg, Germany}
}

\author{Bernd A. Kniehl}{
  address={{II.} Institut f\"ur Theoretische Physik, Universit\"at Hamburg,
Luruper Chaussee 149, 22761 Hamburg, Germany}
}

\begin{abstract}
We report on our recent calculation of the inclusive direct photo- and hadroproduction of the $J/\psi$ meson at next-to-leading order within the factorization formalism of nonrelativistic QCD. We fit the color-octet (CO) long-distance matrix elements $\langle {\cal O}^{J/\psi}(^1\!S_0^{[8]}) \rangle$, $\langle {\cal O}^{J/\psi}(^3\!S_1^{[8]}) \rangle$ and $\langle {\cal O}^{J/\psi}(^3\!P_0^{[8]}) \rangle$ to the transverse momentum ($p_T$) distributions measured by CDF at Fermilab Tevatron and by H1 at DESY HERA and show that they also successfully describe the $p_T$ distributions from PHENIX at BNL RHIC and CMS
at the CERN LHC as well as the photon-proton c.m.\ energy and (with worse
agreement) the inelasticity distributions from H1. In all experiments, the CO processes are shown to be indispensable.
\end{abstract}

\maketitle


\section{Introduction and Overview}

The factorization formalism of nonrelativistic QCD (NRQCD) \cite{Bodwin:1994jh}
provides a rigorous theoretical framework for the description of
heavy-quarkonium production and decay.
This implies a separation of process-dependent short-distance coefficients, to
be calculated perturbatively as expansions in the strong-coupling constant
$\alpha_s$, from supposedly universal long-distance matrix elements
(LDMEs), to be extracted from experiment.
The relative importance of the latter can be estimated by means of velocity
scaling rules; {\it i.e.}, the LDMEs are predicted to scale with a definite
power of the heavy-quark ($Q$) velocity $v$ in the limit $v\ll1$.
In this way, the theoretical predictions are organized as double expansions in
$\alpha_s$ and $v$.
A crucial feature of this formalism is that it takes into account the complete
structure of the $Q\overline{Q}$ Fock space, which is spanned by the states
$n={}^{2S+1}L_J^{[a]}$ with definite spin $S$, orbital angular momentum
$L$, total angular momentum $J$, and color multiplicity $a=1,8$.
In particular, this formalism predicts the existence of color-octet (CO)
processes in nature.
This means that $Q\overline{Q}$ pairs are produced at short distances in
CO states and subsequently evolve into physical, color-singlet (CS) quarkonia
by the nonperturbative emission of soft gluons.
In the limit $v\to0$, the traditional CS model (CSM) is recovered in the case
of $S$-wave quarkonia.
In the case of $J/\psi$ production, the CSM prediction is based just on the
$^3\!S_1^{[1]}$ CS state, while the leading relativistic corrections, of
relative order ${\cal O}(v^4)$, are built up by the $^1\!S_0^{[8]}$,
$^3\!S_1^{[8]}$, and $^3\!P_J^{[8]}$ ($J=0,1,2$) CO states.

The greatest success of NRQCD was that it was able to explain the $J/\psi$
hadroproduction yield at the Fermilab Tevatron \cite{Cho:1995vh}, while the
CSM prediction lies orders of magnitudes below the data, even if the latter
is evaluated at NLO \cite{Campbell:2007ws,Gong:2008ft}. The situation is similar for the transverse momentum ($p_T$) distribution at BNL RHIC \cite{Brodsky:2009cf}.
Also in the case of $J/\psi$ photoproduction at DESY HERA, the CSM cross
section at NLO significantly falls short of the data
\cite{Kramer:1994zi,Butenschoen:2009zy}.
Complete NLO calculations for the CO contributions were performed for inclusive $J/\psi$ production in two-photon collisions \cite{Klasen:2001cu}, $e^+e^-$ annihilation \cite{Zhang:2009ym}, and direct photoproduction \cite{Butenschoen:2009zy}. As for hadroproduction at NLO, before this talk was given, the CO contributions due to intermediate
$^1\!S_0^{[8]}$ and $^3\!S_1^{[8]}$ states \cite{Gong:2008ft} were calculated
as well as the complete NLO corrections to $\chi_J$ production, including both $^3\!P_J^{[1]}$ and $^3\!S_1^{[8]}$ contributions \cite{Ma:2010vd}.

In order to convincingly establish the CO mechanism and the LDME universality,
it had been an urgent task to complete the NLO $J/\psi$ hadroproduction calculation by including the full CO contributions. This was actually achieved in the work \cite{Butenschoen:2010rq} presented at this conference.

Our strategy for testing NRQCD factorization in $J/\psi$ production at NLO is
as follows. We first perform a common fit of the CO LDMEs to the $p_T$ distributions
measured by CDF in hadroproduction at Tevatron Run~II \cite{Acosta:2004yw} and
by H1 in photoproduction at HERA1 \cite{Adloff:2002ex} and HERA2
\cite{Aaron:2010gz} (see Table~\ref{tab:fit} and Fig.~\ref{fig:fitgraphs}).
We then compare the $p_T$ distributions measured by PHENIX at RHIC
\cite{Adare:2009js} and CMS at the LHC \cite{CMSdata} as well as the
$W$ and $z$ distributions measured by H1 at HERA1 \cite{Adloff:2002ex} and
HERA2 \cite{Aaron:2010gz} with our respective NLO predictions based on these CO
LDMEs (see Fig.~\ref{fig:other}). For details on the calculation and the input parameters used, we refer the reader to Ref.~\cite{Butenschoen:2010rq}.

\section{Fit to HERA and Tevatron data}

\begin{table}
\begin{tabular}{@{\hspace{10.3pt}}cc@{\hspace{10.3pt}}}
\hline
$\langle {\cal O}^{J/\psi}(^1\!S_0^{[8]}) \rangle$ &
$(4.76\pm0.71)\times10^{-2}$~GeV$^3$ \\
$\langle {\cal O}^{J/\psi}(^3\!S_1^{[8]}) \rangle$ &
$(2.65\pm0.91)\times10^{-3}$~GeV$^3$ \\
$\langle {\cal O}^{J/\psi}(^3\!P_0^{[8]}) \rangle$ &
$(-1.32\pm0.35)\times10^{-2}$~GeV$^5$ \\
\hline
\end{tabular}
\caption{\label{tab:fit} NLO fit results for the $J/\psi$ CO LDMEs.}
\end{table}

\begin{figure}
\begin{minipage}{7.97cm}
\includegraphics[width=3.95cm]{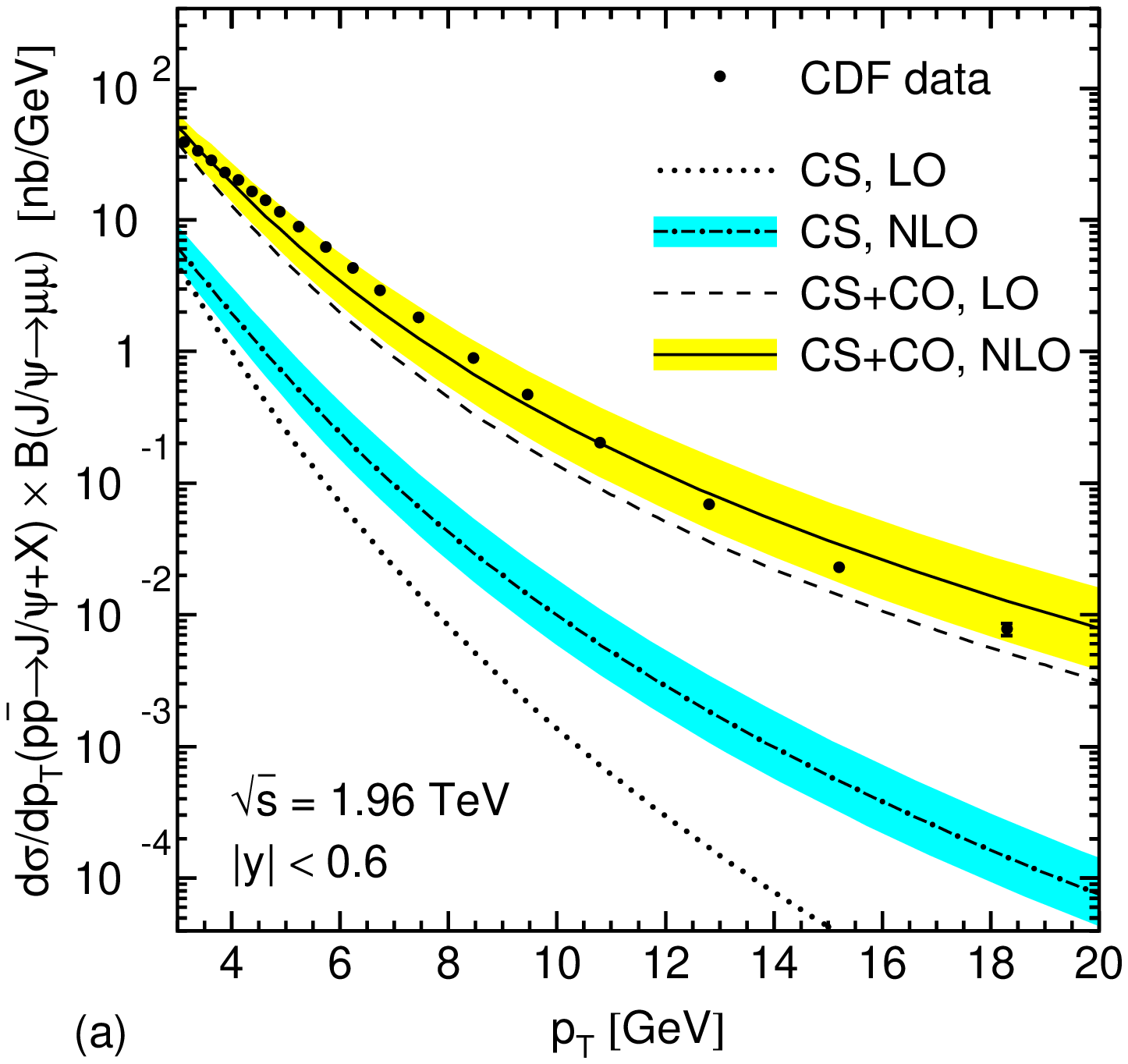}
\includegraphics[width=3.95cm]{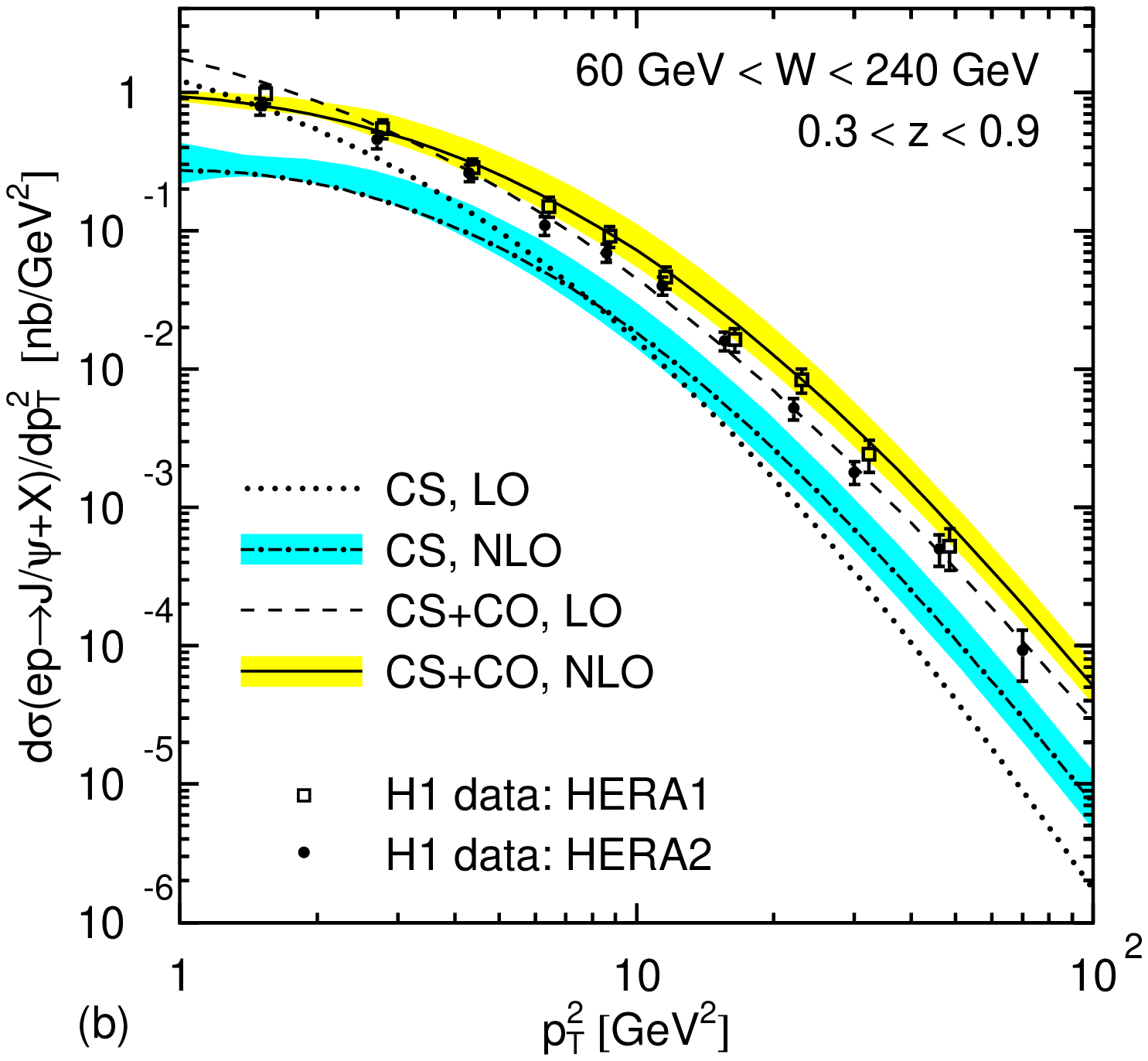}

\vspace{4pt}
\includegraphics[width=3.95cm]{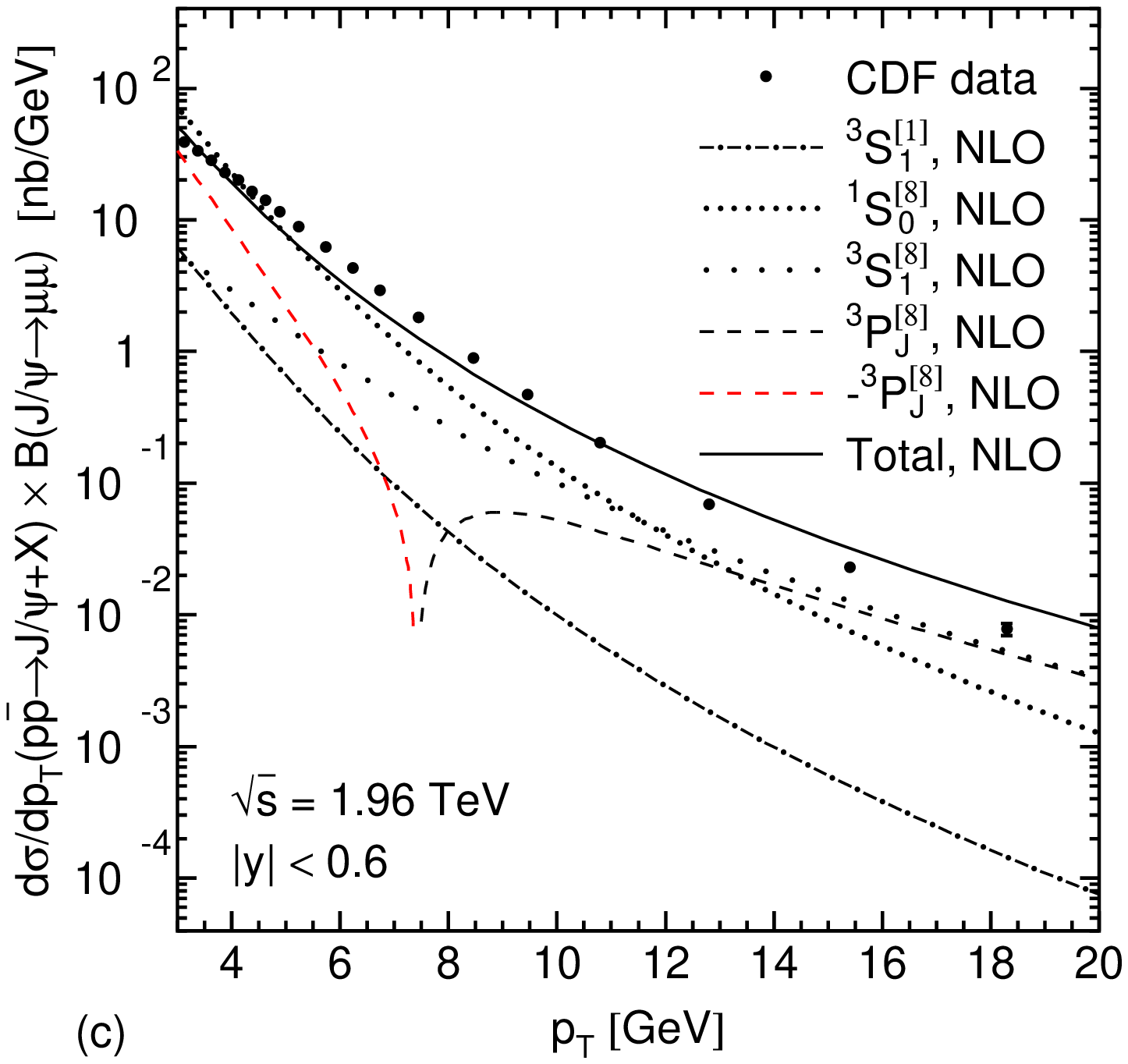}
\includegraphics[width=3.95cm]{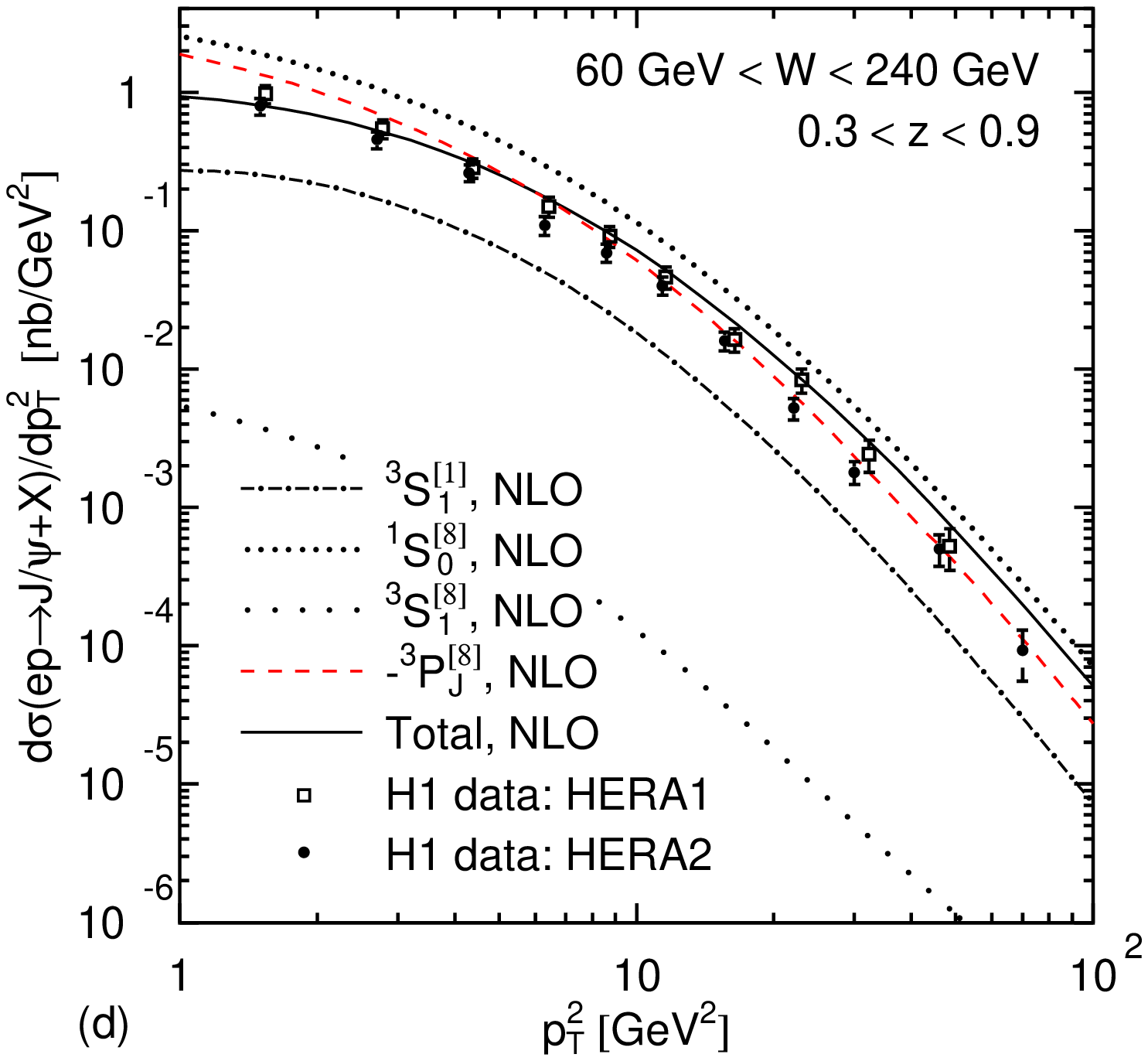}
\end{minipage}
\caption{\label{fig:fitgraphs}
NLO NRQCD predictions of $J/\psi$ hadro- and photoproduction resulting from the
fit compared to the CDF \cite{Acosta:2004yw} and H1
\cite{Adloff:2002ex,Aaron:2010gz} input data.}
\end{figure}

The $p_T$ distribution of $J/\psi$ hadroproduction measured experimentally
flattens at $p_T<3$~GeV due to nonperturbative effects, a feature that cannot
be faithfully described by fixed-order perturbation theory.
We, therefore, exclude the CDF data points with $p_T<3$~GeV from our fit.
We have checked that our fit results depend only feebly on the precise
location of this cutoff.
The fit results for the CO LDMEs corresponding to our default NLO NRQCD
predictions are collected in Table~\ref{tab:fit}.
In Figs.~\ref{fig:fitgraphs}(a) and (b), the latter (solid lines) are compared
with the CDF \cite{Acosta:2004yw} and H1 \cite{Adloff:2002ex,Aaron:2010gz}
data, respectively.
 
For comparison, also the default predictions at LO (dashed lines) as well as
those of the CSM at NLO (dot-dashed lines) and LO (dotted lines) are shown.
In order to visualize the size of the NLO corrections to the hard-scattering
cross sections, the LO predictions are evaluated with the same LDMEs.

We observe from Fig.~\ref{fig:fitgraphs}(c) that the $^3\!P_J^{[8]}$
short-distance cross section of hadroproduction receives sizable NLO
corrections that even turn it negative at $p_T\gtrapprox7$~GeV.
This is, however, not problematic because a particular CO contribution
represents an unphysical quantity depending on the NRQCD scale $\mu_\Lambda$ and the choices of the renormalization scheme and is entitled to become negative as long as
the full cross section remains positive.

In contrast to the situation at LO, the line shapes of the $^1\!S_0^{[8]}$ and
$^3\!P_J^{[8]}$ contributions significantly differ at NLO. Therefore we can now, in our combined HERA-Tevatron fit, independently determine $\langle {\cal O}^{J/\psi}(^1\!S_0^{[8]}) \rangle$ and $\langle {\cal O}^{J/\psi}(^3\!P_0^{[8]}) \rangle$.
Notice that $\langle {\cal O}^{J/\psi}(^3\!P_0^{[8]}) \rangle$ comes out
negative, which is not problematic for the same reasons as explained above for the short distance cross sections.
 
\section{Predictions for further data}

\begin{figure}
\begin{minipage}{7.97cm}
\includegraphics[width=3.95cm]{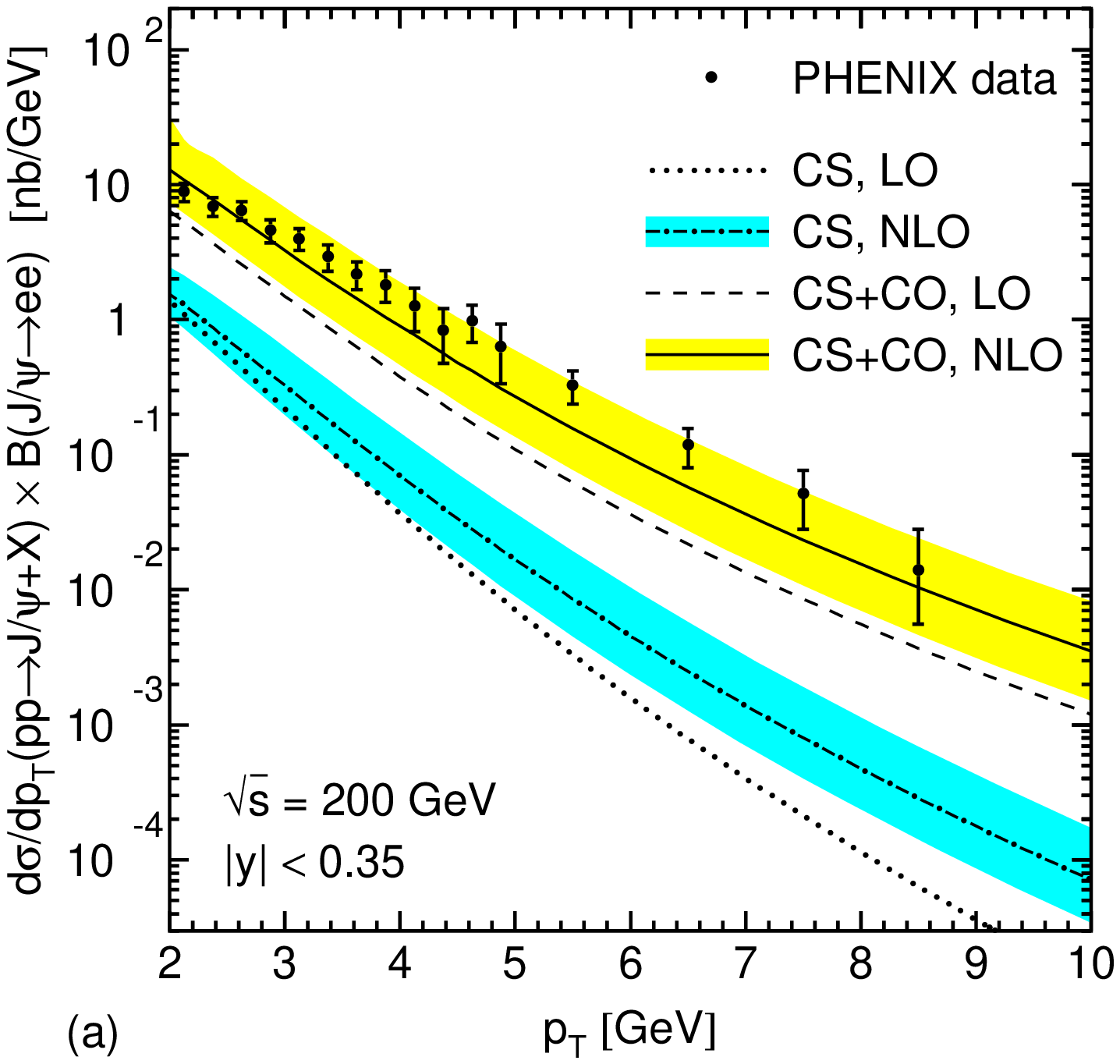}
\includegraphics[width=3.95cm]{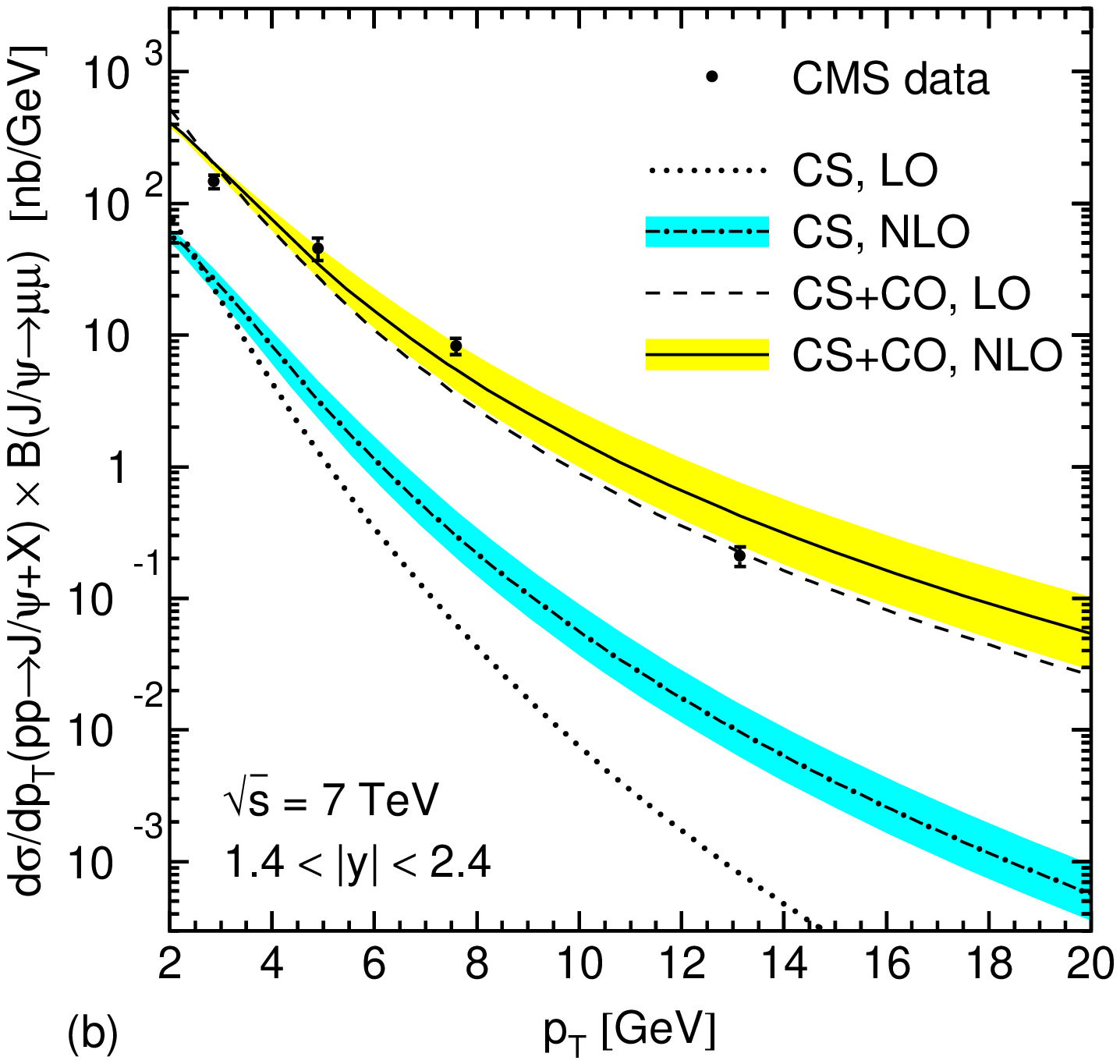}

\vspace{4pt}
\includegraphics[width=3.95cm]{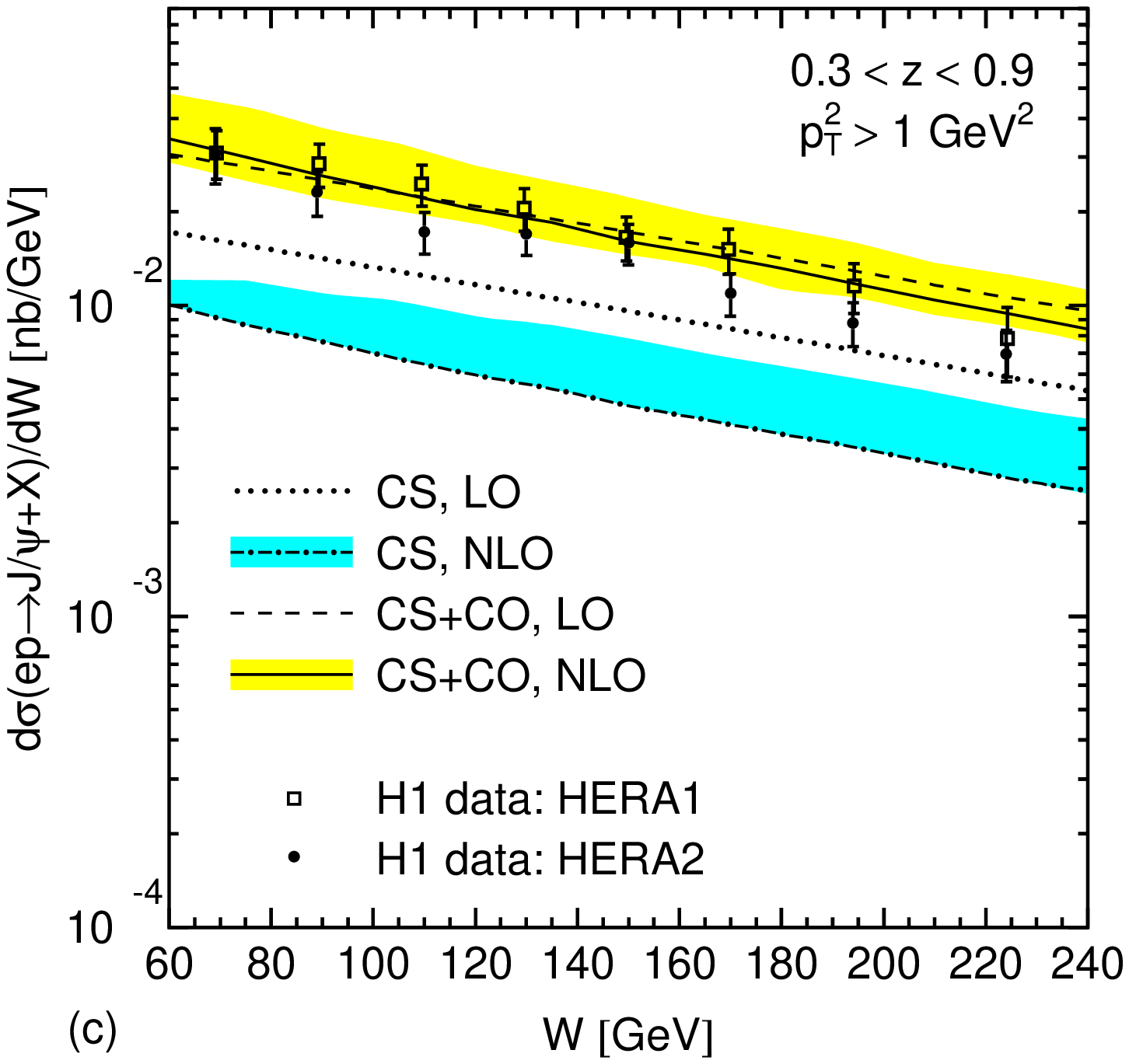}
\includegraphics[width=3.95cm]{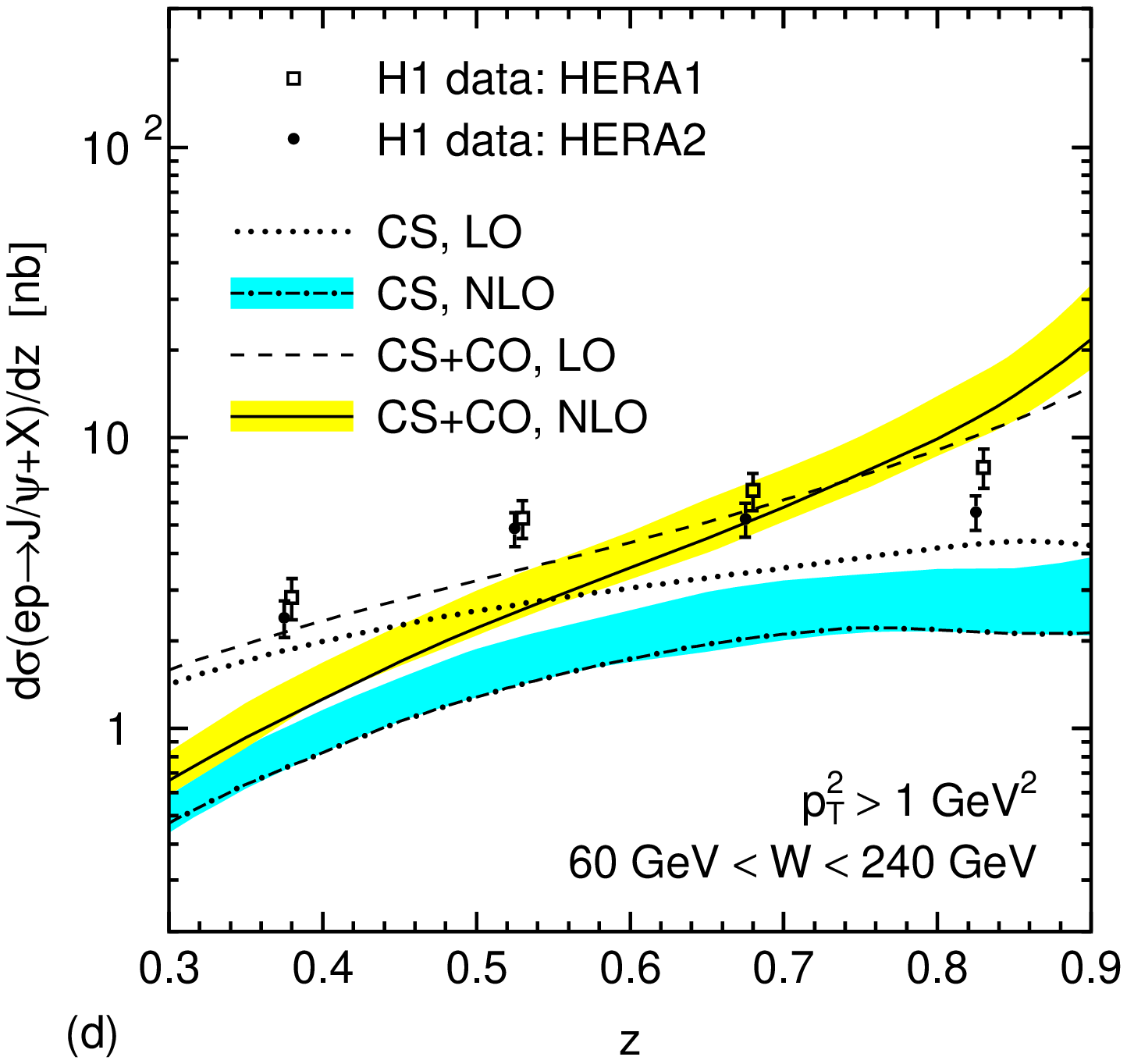}
\end{minipage}
\caption{\label{fig:other}
NLO NRQCD predictions of $J/\psi$ hadro- and photoproduction resulting from the
fit compared to RHIC \cite{Adare:2009js}, CMS \cite{CMSdata}, and H1 
\cite{Adloff:2002ex,Aaron:2010gz} data not included in the fit.}
\end{figure}

We observe from Fig.~\ref{fig:other} that our NLO NRQCD predictions nicely
describe the $p_T$ distributions from PHENIX \cite{Adare:2009js} (a) and CMS
\cite{CMSdata} (b) as well as the $W$ distributions from H1
\cite{Adloff:2002ex,Aaron:2010gz} (c), with most of the data points falling
inside the yellow (light shaded) error band.
The NLO NRQCD prediction of the $z$ distribution (d) agrees with the H1 data
in the intermediate $z$ range, but its slope appears to be somewhat too steep
at first sight.
However, the contribution due to resolved photoproduction, which is not yet
included here, is expected to fill the gap in the low-$z$ range, precisely
where the resolved contribution is peaked.
Near the high-$z$ endpoint region, the NRQCD expansion is understood to break down, and the NRQCD series could be resummed via the introduction of universal shape functions \cite{Beneke:1997qw}, possibly in the context of soft collinear effective theory \cite{Fleming:2006cd}.

\section{Conclusions}

We have performed a complete NLO calculation of the inclusive $J/\psi$ direct photo- and hadroproduction cross sections including the full CS and CO contributions.

We conclude from Figs.~\ref{fig:fitgraphs}(a) and (b) and
\ref{fig:other}(a)--(d) that all experimental data sets considered here
significantly overshoot the NLO CSM predictions, by many experimental standard
deviations. Specifically, the excess amounts to 1--2 orders of magnitude in the
case of hadroproduction and typically a factor of 3 in the case of
photoproduction. On the other hand, these data nicely agree with the NLO NRQCD predictions,
apart from well-understood deviations in the case of the $z$ distribution of
photoproduction.

In our work for the first time a multi-process fit of the CO LDMEs is performed, which come out to be consistent with NRQCD scaling rules. Our work therefore constitutes the most rigorous evidence for the existence of CO processes in nature and the LDME universality since the introduction of the NRQCD factorization formalism 15 years ago \cite{Bodwin:1994jh}.

We should remark that our theoretical predictions refer to direct $J/\psi$
production, while the CDF and CMS data include all prompt events and the H1 and
PHENIX data even non-prompt ones. However, the resulting error turns out to be small against
our theoretical uncertainties and has no effect on our conclusions.

We would like to thank Geoffrey Bodwin and Gustav Kramer for useful discussions and Barbara Jacak and Cesar Luiz da Silva for help with comparing our predictions to the PHENIX data \cite{Adare:2009js}.

\section{Comparison with Ref. \cite{Ma:2010yw}}

As a last point, we compare our results with those obtained by the authors of
Ref.~\cite{Ma:2010yw}, another full NLO NRQCD calculation of $J/\psi$ hadroproduction, which appeared after this talk was held.
Adopting their inputs, we find agreement with their results for the
$^3\!S_1^{[1]}$, $^1\!S_0^{[8]}$, $^3\!S_1^{[8]}$, and $^3\!P_J^{[8]}$
contributions.
However, their fitting philosophy greatly differs from ours.
Specifically, they only fit to Tevatron data with $p_T>7$~GeV, but account for
prompt production, while we jointly fit to Tevatron data with $p_T>3$~GeV and
HERA data neglecting feed-down contributions.
Detailed investigation reveals that the feed-down correction and the shift in
the lower $p_T$ cut on the Tevatron data only moderately affect our joint fit.
However, excluding the HERA data altogether renders the fit greatly
underdetermined.
Faced by this, Ma {\it et al.}\ \cite{Ma:2010yw} perform a constrained fit to
just two linear combinations $M_0$ and $M_1$ of 
$\langle{\cal O}^{J/\psi}(^1\!S_0^{[8]})\rangle$,
$\langle{\cal O}^{J/\psi}(^3\!S_1^{[8]})\rangle$, and
$\langle{\cal O}^{J/\psi}(^3\!P_0^{[8]})\rangle$ defined in such a way that the
${}^3\!P_0^{[8]}$ contribution is effectively accounted for by the
${}^1\!S_0^{[8]}$ and ${}^3\!S_1^{[8]}$ ones multiplied by $M_0$ and $M_1$,
respectively.
Determining the very combinations $M_0$ and $M_1$ from a three-parameter fit
just to the Tevatron data with $p_T>7$~GeV, we find that $M_1$ has attached to
it an error of almost 100\%.
This is because $M_0$ and $M_1$ do not precisely correspond to eigenvectors of
the three-dimensional covariance matrix and the linear combination of
$\langle{\cal O}^{J/\psi}(^1\!S_0^{[8]})\rangle$,
$\langle{\cal O}^{J/\psi}(^3\!S_1^{[8]})\rangle$, and
$\langle{\cal O}^{J/\psi}(^3\!P_0^{[8]})\rangle$ corresponding to the third
eigenvector carries a sizable error feeding into $M_1$.


\end{document}